\documentclass[prd,aps,nofootinbib,superscriptaddress,showkeys]{revtex4}
\usepackage{epsfig}
\usepackage{amsmath}
\usepackage{amsfonts}
\usepackage{amssymb}
\usepackage{color}
\usepackage{graphicx}
\usepackage{graphics}
\usepackage{hyperref}
\usepackage{epstopdf}

\begin{document}

\title{A three-loop radiative neutrino mass model with dark matter}

\author{Li-Gang Jin}
\email{jinligang@njnu.edu.cn}
\author{Rui Tang}
\author{Fei Zhang}
\affiliation{Department of Physics and Institute of Theoretical Physics,
Nanjing Normal University, Nanjing, Jiangsu 210023, People¡¯s Republic of China}

\begin{abstract}
We present a model that generates small neutrino masses at three-loop level due to the existence of Majorana fermionic dark matter, which is stabilized by a $Z_2$ symmetry. The model predicts that the lightest neutrino is massless. We show a prototypical parameter choice allowed by relevant experimental data, which favors the case of normal neutrino mass spectrum and the dark matter with $m\sim 50-135$ GeV and a sizable Yukawa coupling. It means that new particles can be searched for in future $e^+e^-$ collisions.
\end{abstract}

\maketitle

\section{Introduction}
The discovery of very small, but non-zero neutrino
masses and the existence of dark matter (DM) in the Universe may provide important information to guide
us in the search for new physics beyond the standard model (SM). In recent years the idea to incorporate both phenomena in a unified framework has received much attention. And among the simplest realizations is the inert doublet model~\cite{ma,Barbieri,Honorez}, which generates one-loop neutrino masses with the DM being either an extra scalar-doublet or a Majorana fermion whose stability is protected by an exact $Z_2$ symmetry.

Due to the smallness of the neutrino mass scale, a number of models were proposed to generate neutrino masses via higher loop processes, especially via 3-loop ones with the loop suppression $(g^2/16\pi^2)^3\sim 10^{-13}$ ($g$ being a electroweak-sized coupling) to naturally explain the large hierarchy $m_\nu/v \sim 10^{-13}$ ($v$ being electroweak scale). An earlier model~\cite{three-loop krauss} advocated by Krauss, Nasri and Trodden (KNT) extends the SM to include two charged scalar singlets and a right-handed neutrino. Meanwhile, the model has an additional discrete symmetry, which makes neutrino masses be first obtained at the 3-loop level via the new particles with the masses of order of TeV. Therefore, this model is phenomenologically interesting, and is well studied in the subsequent literatures~\cite{Baltz,cheung,Ahriche1,Ahriche2,Ahriche3,Ahriche4,three-loop chen}. Moreover, the generation of 3-loop neutrino masses also appear in the cocktail model~\cite{cocktail}, which adds to the SM two scalar singlets (singly and doubly charged) and a scalar doublet.

In this paper, we present a new model by substituting a scalar triplet with hypercharge $Y=0$ for a charged scalar singlet in the KNT model. Similarly, due to the additional $Z_2$ symmetry and the field content of the model, Majorana neutrino masses are also first generated at the 3-loop level, and the lightest $Z_2$-odd right-handed Majorana fermion could be a DM candidate.

The paper is organized as follows: in Section 2 we describe the model, obtain the neutrino mass matrix, and calculate the DM annihilation processes. Various constraints on the model are analyzed numerically in Section 3. Then conclusions appear in Section 4.

\section{A model for neutrino masses and dark matter}

\subsection{The model}
In addition to SM fields, our model includes several right-handed Majorana fermions $N_{iR}$, a charged $SU(2)_L$ singlet scalar $S^-$ and a triplet scalar $\Delta$ with hypercharge $Y=0$
\begin{equation}
\label{eq:delta}
 \Delta  = \left( {\begin{array}{*{20}{c}}
\frac{1}{\sqrt 2 } \Delta^0  & {\Delta^+ }\\
\Delta^- &  - \frac{1}{\sqrt 2 }\Delta^0
\end{array}} \right) \, .
\end{equation}
The number of $N_{iR}$ will be explained below. Moreover, we introduce a $Z_2$ symmetry under which the new fields are all odd, whereas the SM fields are even. Given the symmetry and particle content of the model, the extra lagrangian will be
\begin{eqnarray}
 \Delta {\cal L}  &=& \frac{1}{2}{\rm Tr}\left[ \left( D_\mu \Delta \right)^2 \right] + \left(D_\mu S^- \right)^\dagger D^\mu S^- + i\overline {N_{iR}} \not \partial N_{iR}\nonumber\\
 &&- \left( \frac{1}{2}m_{N_i} N_{iR}^T C N_{iR} + g_{i\alpha}N_{iR}^T C l_{\alpha R} S^+ + {\rm h.c.}  \right) - V(\Delta ,S^-,\Phi ), \label{eq:lagrangian}
\end{eqnarray}
where $C$ is the matrix of the charge conjugation and the covariant derivatives take the forms
\begin{eqnarray}
  D_\mu \Delta  &=& \partial_\mu \Delta  - i\frac{g}{2}\left[W_\mu ^a \tau ^a,\Delta\right] \, , \\
  D_\mu S^ -  &=& \partial _\mu S^- + ig' B_\mu S^-  \, .
\end{eqnarray}
Here $\tau^a$ $(a=1,2,3)$ is the Pauli matrix. The scalar potential of the new fields and the SM-like doublet $\Phi$ looks like
\begin{eqnarray}
V(\Delta ,{S^ - },\Phi ) &=& -\mu _H^2 \Phi^\dagger \Phi  + \mu_S^2 S^+ S^- + \mu_\Delta^2{\rm Tr}[ \Delta ^2 ] + \lambda_1 (\Phi^\dagger \Phi )^2 + \lambda_2 (S^+ S^- )^2  \nonumber \\
 && + \lambda_3 ({\rm Tr}[\Delta^2] )^2 + \lambda_4(\Phi^\dagger \Phi ) (S^+ S^-) + \lambda_5{\rm Tr}[\Delta^2](S^+ S^- ) \nonumber \\
 && + \lambda_6 \Phi^\dagger \Phi {\rm Tr}[\Delta^2 ] + (\lambda_7 \Phi^\dagger \Delta \widetilde \Phi S^+ + {\rm h.c.} ) \, ,
\end{eqnarray}
where $\tilde \Phi=i\tau^2 \Phi^\dagger$.

As $Z_2$ is exact, $\Delta$ has no vacuum expectation value. After electroweak symmetry breaking, for $\lambda_7 \neq 0$ the charged $Z_2$-odd scalars $\Delta^-$ and $S^-$ will mix
\begin{equation}
  m^2(\Delta^-,S^-) = \left( \begin{array}{*{20}{c}}
2\mu _\Delta ^2 + \lambda _6 v^2 & \frac{\lambda _7}{2} v ^2\\
\frac{\lambda _7}{2} v^2 & \mu_S^2 + \frac{\lambda _4}{2} v^2
\end{array} \right),
\end{equation}
where $v \approx 246$ GeV is the vacuum expectation value of $\Phi$. They will give rise to two charged mass eigenstates
\begin{equation}
  \left( \begin{array}{l}
H_1^ - \\
H_2^ -
\end{array} \right) = \left( {\begin{array}{*{20}{c}}
{\cos \beta }&{\sin \beta }\\
{ - \sin \beta }&{\cos \beta }
\end{array}} \right)\left( \begin{array}{l}
{\Delta ^ - }\\
{S^ - }
\end{array} \right).
\end{equation}
Now the extra scalars are $H_1^-$, $H_2^-$ and $\Delta^0$ with masses 
\begin{eqnarray}
m_{H_1}\leq m_{\Delta^0}=\sqrt{\cos^2\beta m_{H_1}^2+ \sin^2\beta m_{H_2}^2} \leq m_{H_2} \, .\label{eq:massrelation}
\end{eqnarray}

\subsection{Neutrino masses}
\begin{figure}[tbp]
\centering
\includegraphics[width=15cm]{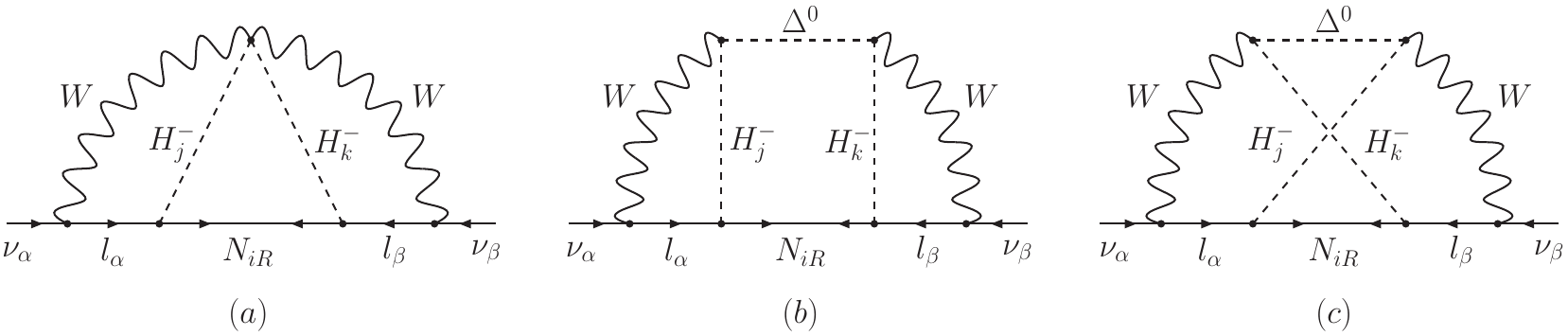}
\caption{\label{Fig:three-loop} Three-loop diagrams for radiative neutrino masses.}
\end{figure}

Explicitly, the lagrangian in Eq.~(\ref{eq:lagrangian}) breaks lepton number, and can generate a Majorana mass for the left-handed neutrinos. However, The $Z_2$ symmetry strictly forbids the generation of neutrino masses at either 1- or 2-loop order, and, therefore, the leading contributions to neutrino masses appear at 3-loop level shown in Fig.~\ref{Fig:three-loop}.

If the model has a single $N_R$, the neutrino mass matrix will predict two vanishing mass eigenvalues like the case in Ref.~\cite{three-loop krauss} and contradict the neutrino oscillation data~\cite{cheung}. In order to solve the problem, one can add small perturbations to the original mass matrix, add more scalars or right-handed Majorana fermions, and so on. In this paper, we employ two right-handed fermions $N_{iR}$ $(i=1,2)$ with $m_{N_{1}}<m_{N_{2}}$, which means that the Yukawa couplings $g_{i\alpha}$ can be complex and bring about three physical CP violation phases. However, in the following discussion, we leave aside the problem of CP violation for simplicity, so $g_{i\alpha}$ takes real number.

For the case of $m_{H_2}>m_{\Delta^0}\gg m_{H_1},m_{N_{i}},m_W$, it is appropriate to neglect the complicated contributions of Figs.~\ref{Fig:three-loop}$(b)$ and \ref{Fig:three-loop}$(c)$. Following the method in~\cite{Ahriche1}, we obtain the neutrino mass matrix elements arising from the remaining Fig.~\ref{Fig:three-loop}$(a)$
\begin{eqnarray}
(M_\nu)_{\alpha\beta} =\sum_{i=1,2}g_{i\alpha}g_{i\beta}m_\alpha m_\beta I_i \, , \label{eq:matrix}
\end{eqnarray}
where $I_i$ is the three-loop integral
\begin{eqnarray}
I_i &=& \frac{g^4 \sin^2(2\beta) m_{N_i}}{6(16\pi^2)^3m_W^4} \int_0^\infty \frac{r \ {\rm d} r}{r +m_{N_i}^2} \times \Big\{ 12\left[F_2(r,m_{H_1}^2,m_{H_2}^2) -F_1(r,m_{H_1}^2,m_{H_2}^2)\right]^2 \nonumber \\ && + \left[G_2(r,m_{H_1}^2,m_{H_2}^2) -G_1(r,m_{H_1}^2,m_{H_2}^2)\right]^2 - F_2(r,m_{H_1}^2,m_{H_2}^2) \big[5 F_2(r,m_{H_1}^2,m_{H_2}^2) \nonumber \\ && -6 F_1(r,m_{H_1}^2,m_{H_2}^2) -G_2(r,m_{H_1}^2,m_{H_2}^2) +G_1(r,m_{H_1}^2,m_{H_2}^2) \big]\Big\}\, . \label{eq:nutrimass}
\end{eqnarray}
Here four integral functions have been introduced 
\begin{eqnarray}
F_1(r,m_{H_1}^2,m_{H_2}^2)&=& \int_0^1 {\rm d}x \ln \frac{x(1-x)r +x m_{H_1}^2}{m_W^2} -(m_{H_1}\rightarrow m_{H_2})\, , \nonumber  \\
F_2(r,m_{H_1}^2,m_{H_2}^2)&=& \int_0^1 {\rm d}x \ln \frac{(1-x)(xr +m_W^2) +x m_{H_1}^2}{m_W^2} -(m_{H_1}\rightarrow m_{H_2})\, , \nonumber  \\
G_1(r,m_{H_1}^2,m_{H_2}^2)&=& \frac{r+m_{H_1}^2}{m_W^2}\int_0^1 {\rm d}x \ x  \ln \frac{x(1-x)r +x m_{H_1}^2}{m_W^2} - (m_{H_1}\rightarrow m_{H_2})\, , \nonumber \\
G_2(r,m_{H_1}^2,m_{H_2}^2)&=& \frac{r-m_W^2+m_{H_1}^2}{m_W^2}\int_0^1 {\rm d}x \ x  \ln \frac{(1-x)(xr +m_W^2) +x m_{H_1}^2}{m_W^2} - (m_{H_1}\rightarrow m_{H_2})\, . \label{eq:nutrimassFG}
\end{eqnarray}

The elements of the neutrino Majorana mass matrix $M_\nu$ can be related to the mass eigenvalues
\begin{eqnarray}
M_\nu= U D_\nu U^T \quad \mbox{with} \quad D_\nu={\rm Diag}(m_1,m_2,m_3)\, ,
\end{eqnarray}
where $U$ is the Pontecorvo-Maki-Nakagawa-Sakata (PMNS) leptonic mixing matrix~\cite{PMNS} parameterized by
\begin{eqnarray}
U= \left ( \begin{array}{ccc} c_{12}c_{13} & s_{12}c_{13} & s_{13}e^{-i\delta} \\ -s_{12}c_{23} -c_{12}s_{23}s_{13} e^{i\delta} & c_{12}c_{23} -s_{12}s_{23}s_{13} e^{i\delta} & s_{23} c_{13} \\ s_{12}s_{23} -c_{12}c_{23}s_{13} e^{i\delta} & -c_{12}s_{23} -s_{12}c_{23}s_{13} e^{i\delta} & c_{23} c_{13}\end{array} \right)\left(\begin{array}{ccc} 1& 0 & 0 \\ 0 & e^{i\alpha_1/2} & 0 \\ 0 & 0 & e^{i\alpha_2/2} \end{array}\right) \label{eq:uu}
\end{eqnarray}
with $c_{ij}=\cos\theta_{ij}$, $s_{ij}=\sin\theta_{ij}$.

Although the numerical results depend on the concrete choice of various parameters in the model, the above neutrino mass matrix has the following special structure
\begin{eqnarray}
M_\nu &=& \left (\begin{array}{ccc} a_{1e} & a_{2e} & 0 \\ a_{1\mu} & a_{2\mu} & 0 \\ a_{1\tau} & a_{2\tau} & 0 \end{array} \right)\left (\begin{array}{ccc} a_{1e} & a_{1\mu} & a_{1\tau} \\ a_{2e} & a_{2\mu} & a_{2\tau} \\ 0 & 0 & 0 \end{array} \right)
\end{eqnarray}
with $a_{i\alpha}=g_{i\alpha}m_\alpha \sqrt{I_i}$. Therefore, the mass of the lightest neutrino is zero for ${\rm Det}(M_\nu)=0$.

\subsection{Dark matter}
When $N_{1}$ is the lightest $Z_2$-odd state, it is stable and can be a WIMP dark matter candidate. For $m_{N_{2}}\gg m_{N_{1}}$, we can safely neglect the effect of $N_{2}$ on $N_{1}$ density. The $N_{1}$ number density get depleted through the annihilation process $N_{1}(p_1)N_{1}(p_2) \rightarrow l_\alpha^+ (p_3) l_\beta^-(p_4)$ via the $t$-channel and $u$-channel exchanges of $H^-_{1,2}$. The amplitude for this process is
\begin{eqnarray}
 {\cal M}_{\alpha\beta} &=&  -g_{1\alpha}^*g_{1\beta}\left( \frac{\sin^2\beta}{t - m_{H_1}^2} + \frac{\cos^2\beta} {t - m_{H_2}^2} \right) \overline u (p_4) P_L u(p_2)\overline v(p_1)P_R v(p_3) \nonumber \\
& & + g_{1\alpha}^* g_{1\beta} \left(\frac{\sin^2\beta}{u - m_{H_1}^2} + \frac{\cos^2\beta}{u - m_{H_2}^2}\right)\overline u (p_4) P_L u(p_1)\overline v(p_2)P_R v(p_3)\, ,
\end{eqnarray}
where $t = (p_1-p_3)^2$ and $u = (p_1-p_4)^2$ are the Mandelstam variables corresponding to the
$t$ and $u$ channels, respectively. After squaring, summing and averaging over the spin
states, the total annihilation cross section in the non-relativistic limit is
given by
\begin{eqnarray}
\sigma v_{\rm rel}&=&\frac{\sum_{(\alpha,\beta)}\left|g_{1\alpha}^* g_{1\beta}\right|^2}{48\pi} m_{N_1}^2v_{\rm rel}^2 \Bigg[\frac{\sin^4 \beta(m_{H_1}^4 +m_{N_1}^4)}{(m_{H_1}^2 +m_{N_1}^2)^4} +\frac{\cos^4 \beta(m_{H_2}^4 +m_{N_1}^4)}{(m_{H_2}^2 +m_{N_1}^2)^4} \nonumber \\
&& + \frac{2\sin^2\beta \cos^2 \beta(m_{H_1}^2m_{H_2}^2 +m_{N_1}^4)}{(m_{H_1}^2 +m_{N_1}^2)^2(m_{H_2}^2 +m_{N_1}^2)^2} \Bigg]\, ,
\end{eqnarray}
where $v_{\rm rel}$ is the relative velocity between the initial particles. Defining $\sigma v_{\rm rel}\equiv a+b v_{\rm rel}^2$, we can approximately relate the dark matter relic abundance to the ¡°$a$¡± and ¡°$b$¡± variables by~\cite{yang bai}
\begin{eqnarray}
\Omega_{N_1}h^2\approx \frac{1.07\times 10^9 \ {\rm GeV}^{-1}}{M_P} \frac{x_F}{\sqrt{g_\star}} \frac{1}{a+3(b-a/4)/x_F}\, ,
\end{eqnarray}
where $M_P= 1.22\times 10^{19}$ GeV is the Planck scale, and $g_\star=86.25$ is the number of relativistic degrees of freedom at the freeze-out temperature $x_F$ given by
\begin{eqnarray}
x_F=\ln \left[\frac{5}{4} \sqrt{\frac{45}{8}} \frac{g}{2\pi^3} \frac{M_P m_{N_1} (a+6b/x_F)}{\sqrt{g_\star x_F}}\right]\, .
\end{eqnarray}
Here $g = 2$ is the number of degrees of freedom for the Majorana fermion dark matter.

\section{Experimental constraints and numerical results}
Firstly, we summarize some relevant experimental data. A global fit to neutrino oscillation data gives~\cite{pdg}
\begin{eqnarray}
s_{12}^2 &=& 0.308\pm 0.017\, , \nonumber \\
s_{23}^2 &=& 0.437^{+0.033}_{-0.023}\ (0.455^{+0.039}_{-0.031})\, , \nonumber \\
s_{13}^2 &=& 0.0234^{+0.0020}_{-0.0019} \ (0.0240^{+0.0019}_{-0.0022})\, , \nonumber \\
\Delta m_{21}^2 &=& 7.54^{+0.26}_{-0.22} \times 10^{-5}\ {\rm eV}^2\, , \nonumber \\
|\Delta m^2| &=& 2.43\pm 0.06\ (2.38\pm0.06) \times 10^{-3}\ {\rm eV}^2\, , \label{eq:neuex}
\end{eqnarray}
where the values (values in brackets) correspond to $m_1 < m_2 < m_3$ ($m_3 < m_1 < m_2$), i.e. normal mass spectrum (inverted mass spectrum), and $\Delta m^2=m_{3}^2-(m_{2}^2+m_1^2)/2$. As mentioned before, the lightest neutrino in the model is massless, thus
\begin{eqnarray}
m_1\simeq 0 \ \ (4.89 \times 10^{-2}) \ {\rm eV} \, , \quad m_2\simeq 8.68\times 10^{-3} \ \ (4.97\times 10^{-2}) \ {\rm eV}\, , \quad m_3 \simeq 4.97\times 10^{-2} \ \ (0) \ {\rm eV} \,.
\end{eqnarray}
The measured value of the relic density from WMAP~\cite{WMAP} and Planck~\cite{Planck} is
\begin{eqnarray}
\Omega h^2=0.1199\pm 0.0027 \, . \label{eq:omega}
\end{eqnarray}

Moreover, the additional $H^-_i$ and $N_{i}$ can mediate 1-loop lepton flavour violating (LFV) processes, such as $l_\alpha\rightarrow l_\beta \gamma$ ($\alpha=\mu,\tau$, $\beta=e,\mu$), and the branching ratios are
\begin{eqnarray}
{\rm Br}(l_\alpha  \to l_\beta \gamma ) &=& \frac{3\alpha}{64\pi G_F^2}\left| \sum\limits_{i = 1,2} g_{i\alpha}^*g_{i\beta }\left[\frac{\sin^2\beta }{m_{H_1}^2} H\left(\frac{m_{N_i}^2}{m_{H_1}^2}\right) + \frac{\cos^2\beta }{m_{H_2}^2}H\left(\frac{m_{N_i}^2}{m_{H_2}^2}\right) \right] \right|^2 \nonumber \\
&\times & {\rm Br}(l_\alpha \to l_\beta \nu_\alpha \bar{\nu}_\beta) \, , \label{eq:brtau}
\end{eqnarray}
where $\alpha=e^2/(4\pi)$ is the electromagnetic fine structure constant, $G_F$ is the Fermi constant, and the function $H(x)$ is given by
\begin{equation}
  H(x) = \frac{1 - 6x + 3x^2 + 2x^3 - 6x^2\ln x}{6(1 - x)^4}\, .
\end{equation}
The current experimental upper bounds of the LFV processes are~\cite{boundmueg,boundtaueg}
\begin{eqnarray}
{\rm Br}(\mu\rightarrow e\gamma)&<& 5.7\times 10^{-13} \, ,  \nonumber \\
{\rm Br}(\tau\rightarrow e\gamma) &<& 3.3 \times 10^{-8}\, ,  \nonumber \\
{\rm Br}(\tau\rightarrow \mu\gamma) &<& 4.4 \times 10^{-8}\, . \label{eq:lch}
\end{eqnarray}

In addition, our model can generate effective four-lepton contact interactions at the 1-loop level, which can be probed in $e^+ e^-$ collisions. Therefore, precise data from LEP will produce limits on the leptophilic dark matter. The detail discussions can be found in Ref.~\cite{LEP limit1,LEP limit2}.

Now, we illustrate the allowed parameter space in the case of normal neutrino mass spectrum. The relevant parameters in the model can be chosen as four particle masses $m_{H_i}$, $m_{N_i}$, a mixing angle $\beta$ and six coupling constants $g_{i\alpha}$.

In general, the structure of the neutrino mass matrix in our model is inclined to the hierarchy of $g_{ie}>g_{i\mu}>g_{i\tau}$, and the observed relic abundance implies that $\sum_{(\alpha,\beta)}|g_{1\alpha}^* g_{1\beta}|$ is of order ${\cal O}(1-10)$ for 50 GeV $<m_{N_1}<$ 200 GeV. Consequently, the Yukawa coupling constants could produce the large LFV branching ratios contradicting the current data, especially for $\mu\rightarrow e\gamma$. However, it is interesting that for suitable parameter values the contributions of $N_1$ and $N_2$ in Eq.~(\ref{eq:brtau}) can cancel out due to the opposite sign between $g_{1e}g_{1\mu}$ and $g_{2e}g_{2\mu}$. 

\begin{figure}[tbp]
\centering
\includegraphics[width=12cm]{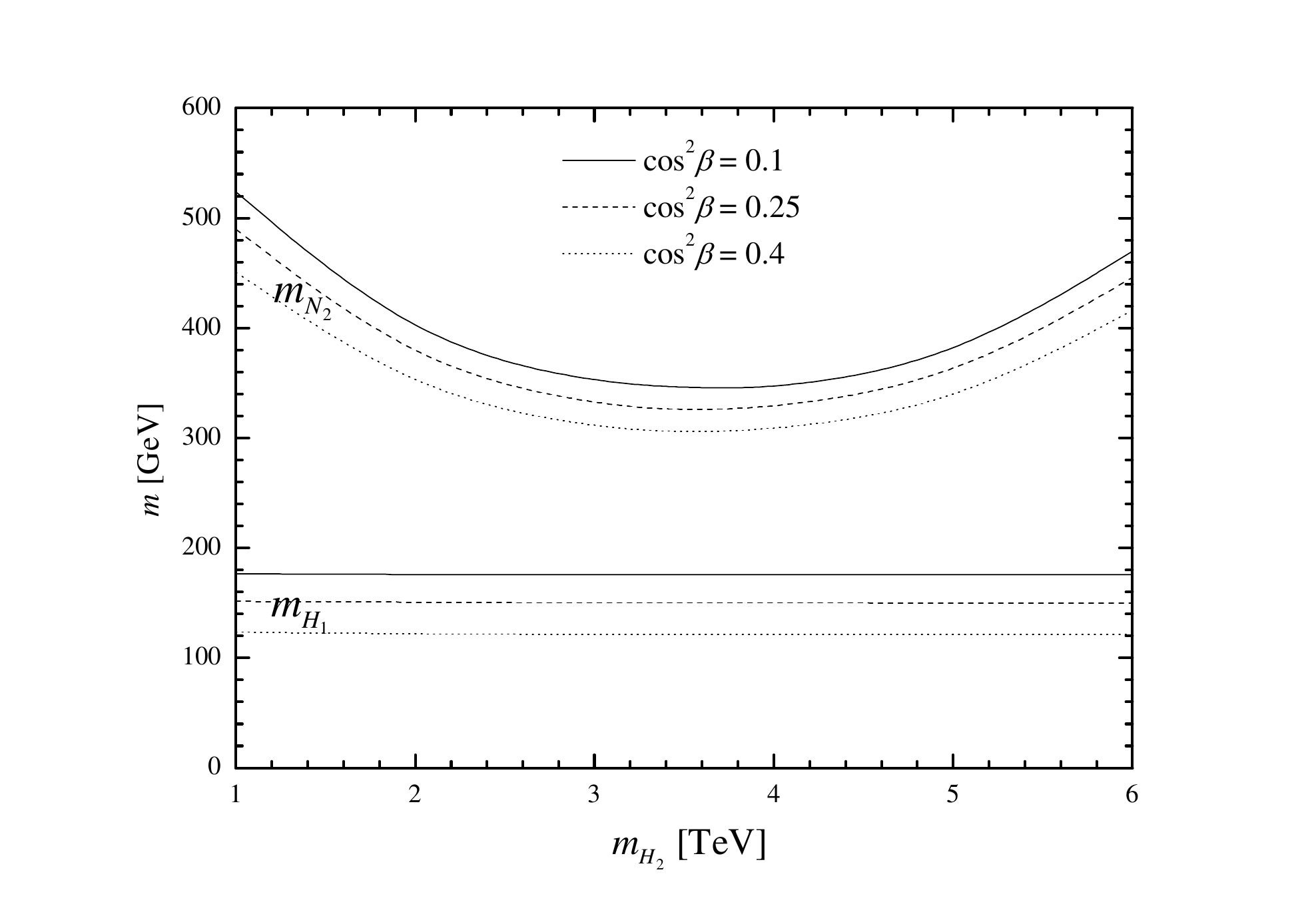}
\caption{\label{Fig:mass} The allowable values of $m_{H_1}$ and $m_{N_2}$ satisfying the neutrino oscillation data, LFV constraints and the observed DM relic density.}
\end{figure}

In Fig.~\ref{Fig:mass}, we keep $m_{N_1}=100$ GeV, $g_{1e}=0.9$, and use experimental data in Eq.~(\ref{eq:neuex}),~(\ref{eq:omega}) and (\ref{eq:lch}) to determine the allowable values of $m_{H_1}$ and $m_{N_2}$ according to $m_{H_2}$ and $\cos^2\beta$. To guarantee the expression of the neutrino mass matrix in Eq.~(\ref{eq:matrix}) only considering the contribution from Fig.~\ref{Fig:three-loop}$(a)$, we pick $m_{H_2}\geq$ 1 TeV, which means that $m_{\Delta^0}$ in Eq.~(\ref{eq:massrelation}) is much larger than $m_{N_1}$, $m_{H_1}$ and $m_W$. In this figure, one can find that the determination of $m_{H_1}$ weakly depends on $m_{H_2}$. For larger $\cos^2\beta$, such as $\cos^2\beta > 0.5$, $N_1$ is heavier than $H_1^-$ and can not be a DM candidate. Meanwhile, for larger $m_{H_2}$, such as $m_{H_2}>7$ TeV, $g_{1e}g_{1\mu}$ and $g_{2e}g_{2\mu}$ have the same sign, which gives rise to unacceptable LFV. 

\begin{figure}[tbp]
\centering
\includegraphics[width=12cm]{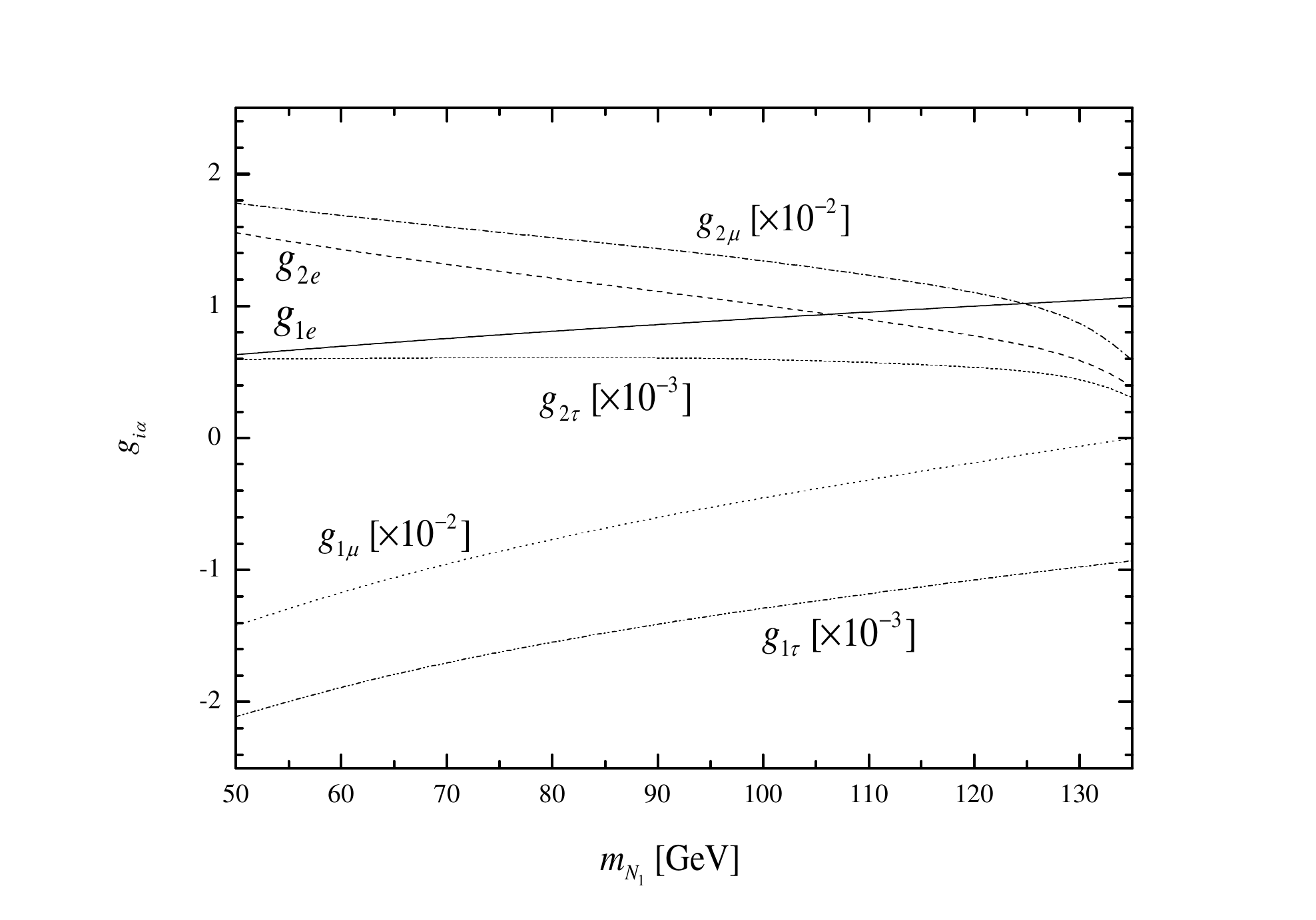}
\caption{\label{Fig:coupling} A prototypical choice of the Yukawa couplings. Note that $g_{i\mu}$ and $g_{i\tau}$ are multiplied by $10^{-2}$ and $10^{-3}$, respectively.}
\end{figure}

In fact, six coupling constants $g_{i\alpha}$ are also definite in the case of Fig.~\ref{Fig:mass}. Now we explicitly present their prototypical values in Fig.~\ref{Fig:coupling}. Here we assume
\begin{eqnarray}
m_{H_2}=5 \ {\rm TeV}\, , \qquad m_{H_1}=1.5m_{N_1} \, , \qquad \cos^2\beta=0.25\, ,
\end{eqnarray}
and $m_{N_2}$ takes a suitable value to realize the cancellation in the decay $\mu\rightarrow e\gamma$. As for $m_{N_1}>135$ GeV, $g_{2\mu}$ is negative, so the cancellation disappears, which gives rise to unacceptable LFV. From the figure, we can give a benchmark in the model. All input parameters are
\begin{eqnarray}
m_{N_1}=100 \ {\rm GeV}\, , \qquad m_{N_2}=350 \ {\rm GeV} \, , \qquad  m_{H_1}=150 \ {\rm GeV}\, , \qquad m_{H_2}=5 \ {\rm TeV}\, ,  \nonumber \\
\cos^2\beta=0.25\, , \quad g_{1e}=0.909\, , \quad g_{1\mu}=-4.52\times 10^{-3} \, , \quad g_{1\tau}=-1.29\times 10^{-3} \, , \nonumber \\
g_{2e}= 1.01 \, , \qquad \qquad g_{2\mu}=1.35\times 10^{-2} \, , \qquad \qquad g_{2\tau}=5.99\times 10^{-4} \, ,
\end{eqnarray}
which leads to the neutrino oscillation data in Eq.~(\ref{eq:neuex}), the DM relic density in Eq.~(\ref{eq:omega}) and the following LFV results
\begin{eqnarray}
{\rm Br}(\mu\rightarrow e\gamma)&=& 5.3\times 10^{-14} \, ,  \nonumber \\
{\rm Br}(\tau\rightarrow e\gamma) &=& 1.8 \times 10^{-12}\, ,  \nonumber \\
{\rm Br}(\tau\rightarrow \mu\gamma) &=& 1.3 \times 10^{-16}\, . 
\end{eqnarray}
They are consistent with the experimental bounds in Eq.~(\ref{eq:lch}).

Moreover, discussion in the case of inverted neutrino mass spectrum is similar, but the bigger coupling constants lead to a much smaller viable parameter space.

\section{Conclusion}
In this paper, we discussed an extension of the SM which includes two right-handed Majorana fermions $N_{i}$, a charged $SU(2)_L$ singlet scalar $S^-$ and a triplet scalar $\Delta$ with hypercharge $Y=0$. Due to the additional $Z_2$ symmetry, the $Z_2$-odd fermion $N_1$ could be a DM candidate and generate Majorana neutrino masses at the 3-loop level. Furthermore, the model predicts that the lightest neutrino is massless for the particular structure of neutrino mass matrix. We also analyzed the constraints on the model coming from relevant experimental data, and presented a prototypical allowed parameter choice, which favors the case of normal neutrino mass spectrum and dark matter with $m\sim 50-135$ GeV and a sizable Yukawa coupling constant $g_{1e}$. It means that the DM and the charged scalar can be searched for in future $e^+e^-$ collisions.

Finally, we did not consider the problem of CP violation in the model. According to recent analyses~\cite{CP phase1,CP phase2}, the best fit value of the Dirac CP violation phase is $\delta\cong 3\pi/2$. Therefore, the coupling constants $g_{i\alpha}$ can be complex, and the model will possess more phenomenology.

\begin{acknowledgments}
This work is supported by the National Natural Science Foundation of China
under Grant No. 11235005.
\end{acknowledgments}


\end{document}